\documentclass[aps,prb,preprint,showpacs,floatfix,superscriptaddress,twocolumn,10pt]{revtex4-2}

\usepackage[english]{babel}
\usepackage{textcomp}
\usepackage{graphicx}
\usepackage{color}
\usepackage{xr}

\usepackage[font=sf]{caption} % Changes font of captions.

\usepackage{amsmath}
\usepackage{amsfonts}
\usepackage{amssymb}
\usepackage{siunitx}
\usepackage{verbatim}
\usepackage{gensymb}
\usepackage{amsthm}
\usepackage{pgfplots}
\usepackage{tikz}
\usepackage{rotating}
\usepackage{xr}
\usepackage{caption}
\usepackage{tabularx}
\usepackage[T1]{fontenc}

\usepackage[]{hyperref}
\begin{document}
\title{First principles characterization of spinterfaces between magnetic Cobaltocene molecule and 2D magnets (CrI$_3$, Fe$_3$GeTe$_2$)}

\author{Nikola Machacova}
\affiliation{Department of Physics and Astronomy, Uppsala University, Box 516, 751\,20 Uppsala, Sweden}

\author{Biplab Sanyal}
\email{For correspondence: Biplab.Sanyal@physics.uu.se}
\affiliation{Department of Physics and Astronomy, Uppsala University, Box 516, 751\,20 Uppsala, Sweden}

\date{\today}

\begin{abstract}
In this paper, we examine the properties of spin-polarized interfaces consisting of single-molecule magnet bis(cyclopentadienyl)cobalt(II) (cobaltocene) and two-dimensional magnetic  materials, semiconducting CrI$_3$ and metallic Fe$_3$GeTe$_2$, using first-principles density functional theory based calculations. Our calculated adsorption energies indicate the stability of these hetero-interfaces with the observation of hybridization of electronic states across the interface. Magnetic properties are characterized using Heisenberg exchange parameters obtained from both total-energy differences and the Liechtenstein-Katsnelson-Antropov-Gubanov (LKAG) formalism in the basis of maximally localized Wannier functions (MLWFs). Analysis of these parameters shows a strong directional anisotropy in the magnetic substrate–molecule interaction in agreement with the nature of orbital hybridization. Additionally, possible exchange mechanisms are proposed based on orbital-resolved exchange and hopping parameters. We also show that the molecular adsorption may enhance the intralayer exchange interactions, with some exchange parameters reaching up to a~threefold increase in magnitude compared to the freestanding case. Finally, we observe a 100~\% spin polarization at the Fermi level in the cobaltocene/CrI$_3$ interface, which makes it particularly promising for spin-transport applications.
\end{abstract}

\maketitle
%%%%%%%%%%%%%%%%%%%%%%
\section{Introduction}
\label{sec:intro}

%broad context about the field - nanoelectronics, spintronics  %why this topic is relevant nowadays

The discovery of giant magnetoresistance effect~\cite{baibich1988giant} sparked enormous interest in spin-dependent phenomena that could be exploited in practical applications. Nowadays, promising alternative to current charge-based devices lies in spintronics~\cite{zhang2021recent}, which exploits the electron's intrinsic magnetic moment - spin - as an~information carrier, which can be effectively manipulated with electric and magnetic fields. While spintronic devices can in principle become faster and more energy-efficient than conventional electronics, practical spintronic systems still face challenges such as low spin injection efficiency and short spin diffusion lengths~\cite{gu2023challenges,natterer2017reading}. Moreover, the miniaturization of devices may be associated with low switching barriers, which increase the chance of unwanted switching of quantum states due to thermal noise~\cite{cavin2006research}. The successful synthesis of graphene has opened an entirely new route of realizing atomically thin materials, characterized as 2D materials with extraordinary properties tunable by chemical functionalization, strain and external fields. In recent times, in the realm of spintronics, several magnetic tunnel junctions (MTJ) based on 2D materials, such as CrI$_3$~\cite{song2018giant, yan2020significant} or Fe$_3$GeTe$_2$ (FGT)~\cite{lin2020ultrathin, zhang2020perfect}, have been proposed in combination with non-magnetic 2D materials - graphene, BN, or InSe.

%molecular magnets
Single molecule magnets \cite{christou2000single} have attracted a lot of interest owing to their potential applications, ranging from quantum computing ~\cite{troiani2011molecular} to high density data storage ~\cite{sessoli2017magnetic}. Molecular systems are promising since they are highly tunable via chemical functionalization, and can support longer spin lifetimes owing to weak spin-orbit coupling~\cite{sanvito2011molecular, gu2023challenges}, but their magnetic properties on surfaces might be altered or even vanish~\cite{holmberg2015adhering}. A~simple example of molecule with single magnetic center are metallocenes~\cite{haaland1979molecular} and their derivatives, commonly used in catalysis of polyolefins~\cite{brintzinger1995stereospecific}.
Furthermore, they could find use in quantum circuits with interesting transport properties~\cite{tang2025unusual}, medicinal chemistry~\cite{santos2017recent}, and capacitors~\cite{mao2013metallocene}.

Recently, metallocenes have been used in scanning tunneling microscopy (STM) for functionalization of the tip retaining their magnetic properties~\cite{ormaza2017efficient}. Using cobaltocene or nickelocene molecule~\cite{garnier2020kondo, soler2025magnetic, bae2025direct, czap2019probing} presents an alternative to spin polarized STM, which is often limited by its reactivity~\cite{soler2025magnetic}.
The approach utilizes Kondo resonance~\cite{scott2010kondo} and allows for determination of effective exchange interaction. This approach has been recently applied to 2D materials and thin films, which present more complicated scenario than single magnetic adatom on surface, probing local exchange fields \cite{aguirre2024ferromagnetic, fetida2024single}. Moreover, previous work has also shown that surface-molecule tunnel junction for high-performance spin-based devices can provide nearly 100~\% polarized spin current \cite{bae2025direct}.
 
The interaction between organometallic compounds and metallic substrates or non-magnetic 2D material has been extensively studied \cite{li2014single,cervetti2016classical,oppeneer2009nature, campbell2016engineering}. However, when it comes to interfaces of 2D magnetic materials, the knowledge is limited to simple and non-magnetic molecules~\cite{tang2020tunable,he2019remarkably, rivero2024magnon}. Only recently there has been a~work on molecular qubits placed on 2D vdW ferromagnet~\cite{dey2025coupling}. Unlike metallic surfaces, 2D ferromagnets are fundamentally different in nature, often exhibit complex physical phenomena and their properties can be tuned via the number of layers, stacking or interlayer magnetic coupling. 

%expected outcome
Upon adsorption, both the molecule and the substrate are expected to influence each other. Their interaction may result in changes in geometry, charge transfer, or magnetic order and anisotropy, which could potentially lead to modified exchange interactions and the emergence of non-collinear magnetic order. In general, the interaction might be slightly weaker than that with metallic substrates, enabling preservation of unique properties of the molecule.

% research question
This work aims to deepen the understanding of the interfaces of magnetic molecule and 2D magnets mainly focusing on charge transfer, electronic structure, and exchange interactions by first principles density functional theory based calculations. The paper is organized as the following. In the next section, we describe the computational methods. In the section 'Results and Discussions', we describe the electronic structure characterization of the isolated molecule followed by the structural properties and stability of the heterointerfaces. The next subsection covers the description of electronic structure followed by the subsection on magnetic properties where we discuss in detail the exchange interaction parameters along with possible exchange pathways. Finally, we present some results on constrained moment calculations for the simulation of the effects of external stimuli, followed by the concluding remarks.
%how can the knowledge be used
%The knowledge might find use in the design of new spintronic devices and the magnetism manipulation at the atomic scale. Moreover, the precision of existing sensing methods could be increased owing to more complex calculations of the interfaces.

%%%%%%%%%%%%%%%%%%%%%%
\section{Computational Methods}
\label{sec:methods}

Electronic structure calculations were performed using the Vienna Ab initio Simulation Package (\textsc{VASP}) \cite{kresse1996efficient} utilizing Projector Augmented Wave potentials \cite{blochl1994projector,kresse1999ultrasoft}. The Perdew-Burke-Ernzerhof (PBE) \cite{perdew1996generalized} exchange-correlation functional was employed at the GGA level. Cutoff energy of the plane waves was set to 500 eV with the electronic convergence criterion as $1\cdot 10^{-7}$~eV for electronic structure calculations. For the calculations of magnetocrystalline anisotropy energy (MAE) with the inclusion of spin-orbit interaction, the precision was increased to $1\cdot 10^{-8}$~eV. The Brillouin zone was sampled with a$\Gamma$-centered $k$-point grid. For the CoCp$_2$/CrI$_3$ system, a 5$\times$5$\times$1 k-point grid was used in a 2$\times$2 supercell of CrI$_3$ for the electronic structure calculations and the determination of MAE. For the FGT-based system, the k-point meshes were 15$\times$15$\times$1 for MAE calculations, 9$\times$9$\times$1 for the electronic structure calculations, and 5$\times$5$\times$1 for the conversion to Wannier functions. In the self-consistent noncollinear calculations, the k-point grids were reduced to 3$\times$3$\times$1 for CrI$_3$ and 7$\times$7$\times$1 for FGT interfaces.

Vacuum of at least 15~\AA{} was included in the unit cells to eliminate interactions along $z$-axis due to periodic boundary conditions. The geometry optimization criterion was set to 0.01~eV/\AA{} for the interfaces and 5~meV/\AA{} for freestanding materials. To account for weak van der Waals interaction, the DFT-D3 correction by Grimme \textit{et al.}~\cite{grimme2010consistent} was included. DFT+U calculations were carried out within the framework of the Dudarev's approach as implemented in VASP~\cite{shishkin2019dftu} to account for electron correlations in $d$-orbitals of Cr and Co ($U_\text{eff}$=3~eV and 4~eV respectively). 
Output data were processed using the \textsc{vaspkit} tool~\cite{wang2021vaspkit}.

The basis change from plane waves to maximally localized Wannier functions (MLWF) was implemented by the \textsc{wannier90} software~\cite{mostofi2008wannier90} with initial projections from the VASP2Wannier interface. The projections included Co $d$-, C $p$-, and H $s$-states from the molecule, Cr and Fe $d$- states for the magnetic atoms, and additionally $p$-states of I, Ge, Te atoms to capture relevant bonding character. The convergence tolerance for wannierization was set to $1 \cdot 10^{-5}$ \AA$^2$. Subsequently, the resulting Hamiltonian in MLWF basis was used to extract magnetic exchange parameters with the \textsc{TB2J} code \cite{he2021tb2j}, based on the Liechtenstein–Katsnelson–Antropov–Gubanov (LKAG) formalism \cite{liechtenstein1987local}. The Heisenberg Hamiltonian used in the calculation reads
\begin{equation}
    H= - \sum_{i,j} J_{ij} \,\vec{e}_i \cdot \vec{e}_j
\end{equation}

Calculations with constrained local magnetic moment were performed employing the penalty functional approach as implemented in \textsc{VASP} \cite{ma2015constrained}, where only the direction of the magnetic moment was fixed. The total energy of the system is given by:
\begin{equation}
    E=E_0+\sum_I\lambda\left[ \Vec{M}_I - \hat{M}_I^0(\hat{M}_I^0 \cdot \Vec{M}_I) \right]^2
\end{equation}
where E$_0$ denotes standard self-consistent DFT energy and $\hat{M}_I^0$ is a~unit vector in the desired direction of the magnetic moment for a given site $I$. The local magnetic moment was obtained via the integration inside a~sphere of Wigner-Seitz radius centered at a~given atomic site. The radii were chosen based on the Bader charge analysis. Since the energy originating from the penalty functional is inversely proportional to $\lambda$ and vanishes in the limit of $\lambda \rightarrow \infty$ \cite{ma2015constrained}, the $\lambda$ parameter was gradually increased until the value of the energy penalty function was smaller than $5\cdot 10^{-8}$~eV per unit cell. 
Monte Carlo spin dynamics simulations were performed by UppASD code~\cite{eriksson2017atomistic, skubic2008method} using Metropolis algorithm with 100 000
steps and 3 ensembles. Several system sizes were tested and periodic boundary conditions were set within the atomic plane.

%%%%%%%%%%%%%%%%%%%%%%
\section{Results and Discussions}
\label{results}
To investigate the interaction between the magnetic molecule and the low-dimensional substrate, two 2D magnetic materials with distinct properties, semiconducting CrI$_3$ and metallic Fe$_3$GeTe$_2$, were chosen as substrates. Bis(cyclopentadienyl)cobalt (cobaltocene, CoCp$_2$) was considered as the magnetic molecule with one unpaired electron.

\subsection{Cobaltocene}
The freestanding cobaltocene molecule was first characterized to provide a~reference. Cobaltocene consists of a~metal centre (Co) and two cyclopentadienyl (Cp) rings. It is a~structural analogue of ferrocene and exists in two conformations with different alignment of the carbon rings, eclipsed conformation (D$_{5h}$ point group) and staggered conformation (D$_{5d}$ group) (Figure S1). %\ref{SI-Cobaltocene-configuration}
Experimental measurements suggest that the rotation energy barrier is relatively small and the outcome is strongly dependent on conditions~\cite{antipin1993redetermination}.
In accordance with previous theoretical studies~\cite{xu2003systematic,adhalsteinsson2023ionization}, eclipsed configuration was found to be more stable (by 20.6~meV at the GGA level) and was used in this work. In our calculations, the high symmetry structure with D$_{5h}$ symmetry changes to C$_{2v}$ due to Jahn-Teller effect. The magnetic Co atom has 3d$^7$ electronic configuration with electronic levels split due to the presence of the ligand field (Figure S2)
%\ref{SI_Cobaltocene_elstr}
)
and one unpaired electron carrying the magnetic moment.

Supercell calculations, which are necessary later in the study, make it impossible to rely on advanced approaches for the description of correlated $d$--electrons (such as hybrid functionals, or DMFT) which are computationally highly demanding. GGA+U approach is therefore employed in the study using $U_\text{eff}$=4~eV for Co atom in cobaltocene. This $U_\text{eff}$ value was selected owing to a~close match with the results of hybrid functional (B3LYP) calculations for a~freestanding molecule, mainly regarding the hybridization of occupied electronic states. Moreover, the spin magnetic moment per molecule necessarily needs to correspond to the one unpaired electron found in the system. Detailed information on the determination of the Hubbard parameter can be found in the SI (Table S1).%~\ref{SI-hubbard-table}

\subsection{Structural properties}
The properties of the interfaces were modeled in a $2\times2$ supercell of the substrate as it provides a~reasonable balance between computational cost and accuracy of the model. Several sizes of the substrate cell were tested. It was found that increasing the size of computational cells in general results in lower moment on Co, with significantly slower convergence for CoCp$_2$/CrI$_3$ interface. The properties of the junction are then dependent on molecule coverage on the surface as higher cobaltocene concentration results in smaller delocalization of electrons into the monolayer. The details on convergence can be found in SI (Tables S2 and S3%SI tables on magnetic moments
).
The orientation of the molecule on the surface was considered with the main rotation axis (C$_5$ in high-symmetry conformations) either parallel or perpendicular to the surface.
Only the structures with the molecule in the perpendicular orientation converged properly and were used in this study. To identify the most stable adsorption configuration, the molecule was systematically translated in the $xy$-plane by creating a 4$\times$4 spatial grid. After structural relaxations, the energies were compared and the differences were typically on the order of tens of meV (Fig.~S6%\ref{} SI figure translation maps
). 

\begin{figure}
    \centering
\includegraphics[width=0.98\linewidth]{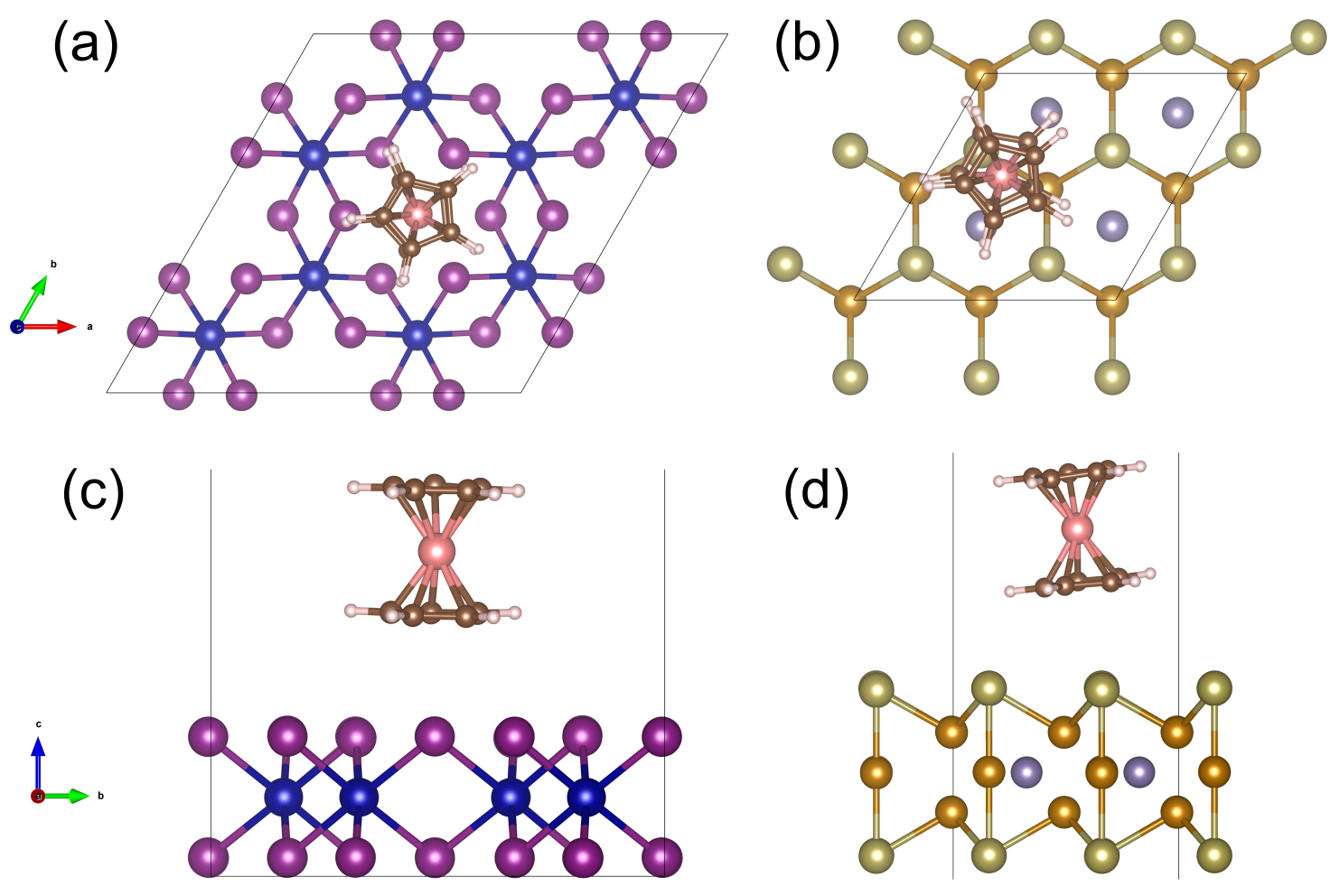}
\caption[Final structures of the interfaces after optimization]{Final structures after optimization. Top view and side view of CrI$_3$ (a,c), Fe$_3$GeTe$_2$ (b,d) interfaces with cobaltocene. The atoms' colors are Cr (blue), I (purple), Fe (orange), Ge (light purple), Te (green), Co (pink), C (brown), and H (white).}
    \label{fig1}
\end{figure}

The average distance between cobaltocene and 2D magnets was similar in both cases, 3.26~\AA{} for CrI$_3$, and 3.25~\AA{} for FGT, which can be considered as the typical value of van der Waals distance. However, it should be noted that our value is larger than previously reported value for molecule-metal interface (2.59~\AA{} in CoCp$_2$/Cu(100)~\cite{garnier2020kondo}).
While the carbon rings were parallel to the CrI$_3$ surface, in the FGT case, the molecule became slightly slanted probably due to the surface electron density being highly anisotropic (Fig.~\ref{fig1}). More structural information is summarized in Table~\ref{tab:vdw-distance}.

%adsorption energy
To assess the stability of the molecule-substrate interfaces, adsorption energies were determined at the GGA level accounting for the dispersion forces. In this work, the adsorption energy is defined as the energy difference between the total energy of the interface and the energies of freestanding substrate and molecule: \begin{equation}
     E_{ads}=  E_{tot}-E_{sub}-E_{mol}.
\end{equation} The resulting values were \mbox{-0.96}~eV for CoCp$_2$/CrI$_3$, and \mbox{-0.63}~eV for CoCp$_2$/FGT in 2$\times$2 cell, so both interfaces are expected to be stable.

\begin{table}[htp]
\centering
\caption[Structural parameters of interfaces]{Unit cell lattice parameters $a$, $b$, average ($d_{\text{avg}}$) and minimal ($d_{\text{min}}$) van der Waals distances (in \AA) and adsorption energy (eV/supercell) for the interfaces in 2$\times$2 supercells, determined using GGA+U method.}
\label{tab:vdw-distance}
%\begin{tabularx}{0.45\textwidth}{xxxxx}
\begin{tabularx}{0.35\textwidth}{c|c|c|c|c}

\hline
Substrate & $a$ (\AA) & $b$ (\AA) & \(d_{\text{avg}}\) (\AA) & \(d_{\text{min}}\) (\AA) % & $E_{ads}$ 
%Substrate & $a$ (\AA) & $b$ (\AA) & \(d_{\text{avg}}\) (\AA) & \(d_{\text{min}}\) (\AA) & $E_{ads}$ (eV/supercell)
\\
\hline
CrI$_3$ & 13.88 & 13.89 & 3.26 & 3.16 %& -0.96
\\
Fe$_3$GeTe$_2$ & 8.03 & 8.02 & 3.25 & 2.97 %& -0.63
\\
\hline
\end{tabularx}
\end{table}

\subsection{Charge transfer}

\begin{figure}[htp]
\centering

\includegraphics[width=0.95\linewidth]{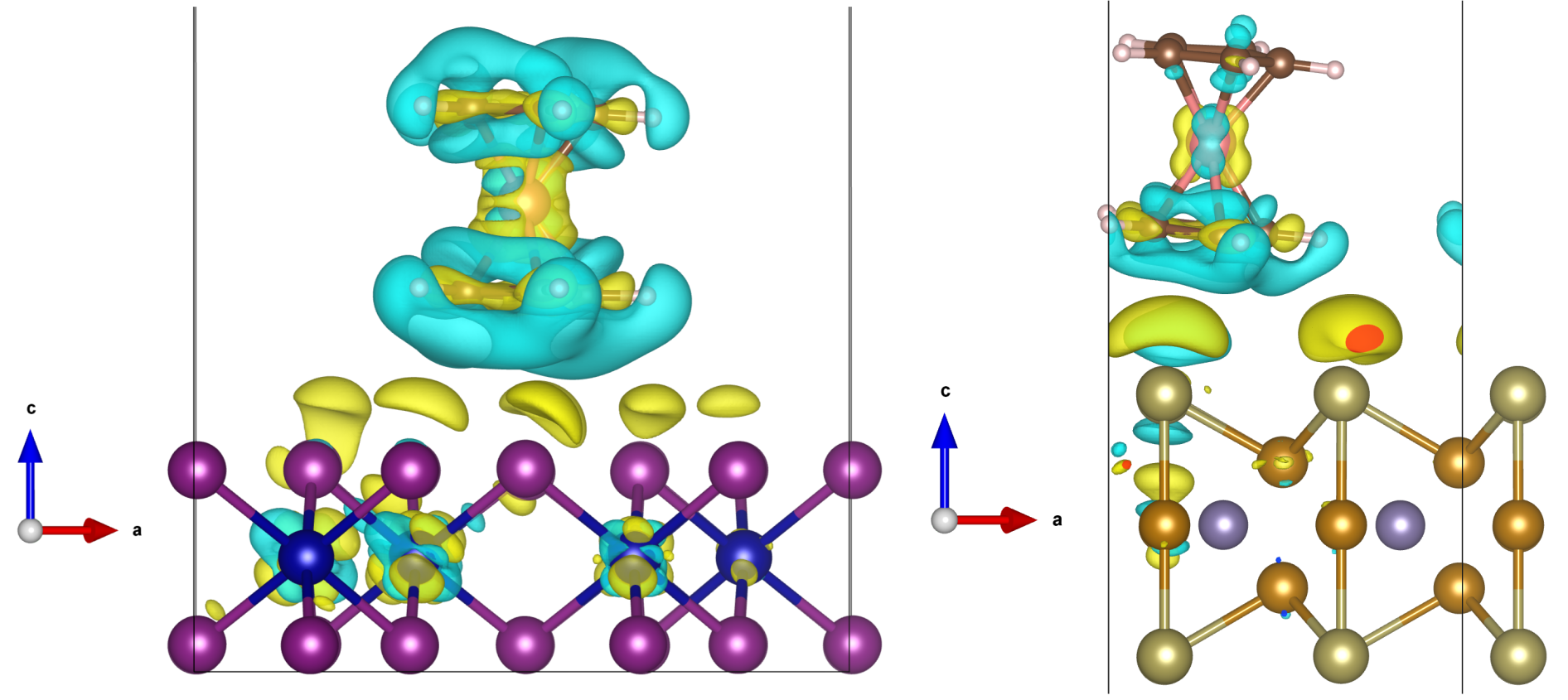} \caption{Charge density differences between the deformed layers and the interface for CrI$_3$ (left), and Fe$_3$GeTe$_2$ (right). The corresponding isosurface levels are: 
$5.9 \cdot 10^{-4} \,e$ (CrI$_3$),
$2.9 \cdot 10^{-4} \,e$ (Fe$_3$GeTe$_2$). }
    \label{fig2}
\end{figure}

%Charge density difference
The surface-molecule interaction is complex, and in general, can also involve charge transfer. Significant changes in electron density were observed at the CoCp$_2$/CrI$_3$ interface with 0.47~$e$ transferred to the substrate as predicted by the Bader charge analysis. The additional charge was redistributed mainly between atoms in the top iodine atomic layer. In contrast, for FGT, the transfer of only 0.03~$e$ was observed. 
%add charge density
To resolve the modifications to electronic charge due to molecular adsorption, we calculated the charge density difference ($\Delta \rho$) as:
\begin{equation} \Delta\rho = \rho_{\text{interface}} -(\rho_{\text{molecule}} + \rho_{\text{substrate}}). \end{equation}
where $\rho_{\text{interface}}$, $\rho_{\text{molecule}}$, and $\rho_{\text{substrate}}$ represent the charge densities of the whole system and the isolated subsystems, respectively. For the subsystem calculations, the atomic coordinates were kept fixed at their optimized positions within the interface. The resulting redistribution of charge density is visualized in Fig.~\ref{fig2}.
The cobaltocene molecule is a 19-electron complex and according to the 18-electron rule, it is therefore expected to be an~electron donor. The change of the electron density could therefore be related to the stabilization of the molecule by transferring a portion of the charge density onto the substrate atoms. Furthermore, it seems that the behavior of the substrate may facilitate the charge transfer, as the electronegativity of iodine is higher than that of both carbon and tellurium and therefore, a~large transfer is observed at CrI$_3$-based interface.
%{\bf Is there any role of electronegativity of I ?}

%Work function
It is important to assess the change in the work function of the substrate due to the molecular adsorption. Work function is a~fundamental characteristic feature of surfaces defined as an~energy necessary for removing an~electron from the material into vacuum:
\begin{equation}
    \Phi= -e \phi - E_F,
\end{equation}
where $e$ is the elementary charge, $\phi$ is the electrostatic potential in vacuum, and $E_F$ is the Fermi level.
For Fe$_3$GeTe$_2$, the work function remained nearly unchanged with a value of 4.12~eV compared to
4.05~eV obtained for the freestanding layer. A~significant decrease in the work function was found for the semiconducting substrate CrI$_3$ where it decreased from 5.65~eV to 4.79~eV. The line profiles of the electrostatic potential can be found in the SI (Figure S7). 

%\begin{figure}
%\centering
%\includegraphics[width=1\linewidth]{Figures/workfunction.png}
%\caption{Electrostatic potential as a~function of distance along the $z$-axis for interfaces with CrI$_3$ (top), Fe$_3$GeTe$_2$ (bottom). The curves for freestanding substrate (blue) and for the system with adsorbed molecule (red) were aligned on the $x$ axis for a~better comparison. The atomic structures are provided for reference.}
%    \label{fig:workfunction}
%\end{figure}

\subsection{Electronic properties}
%electronic
Freestanding substrates were first characterized for a~reference (Figure S3).
Afterwards, both interfaces were examined for changes in electronic structure due to molecule adsorption. 

\begin{figure*}[htp]
    \centering
\includegraphics[width=0.95\textwidth]{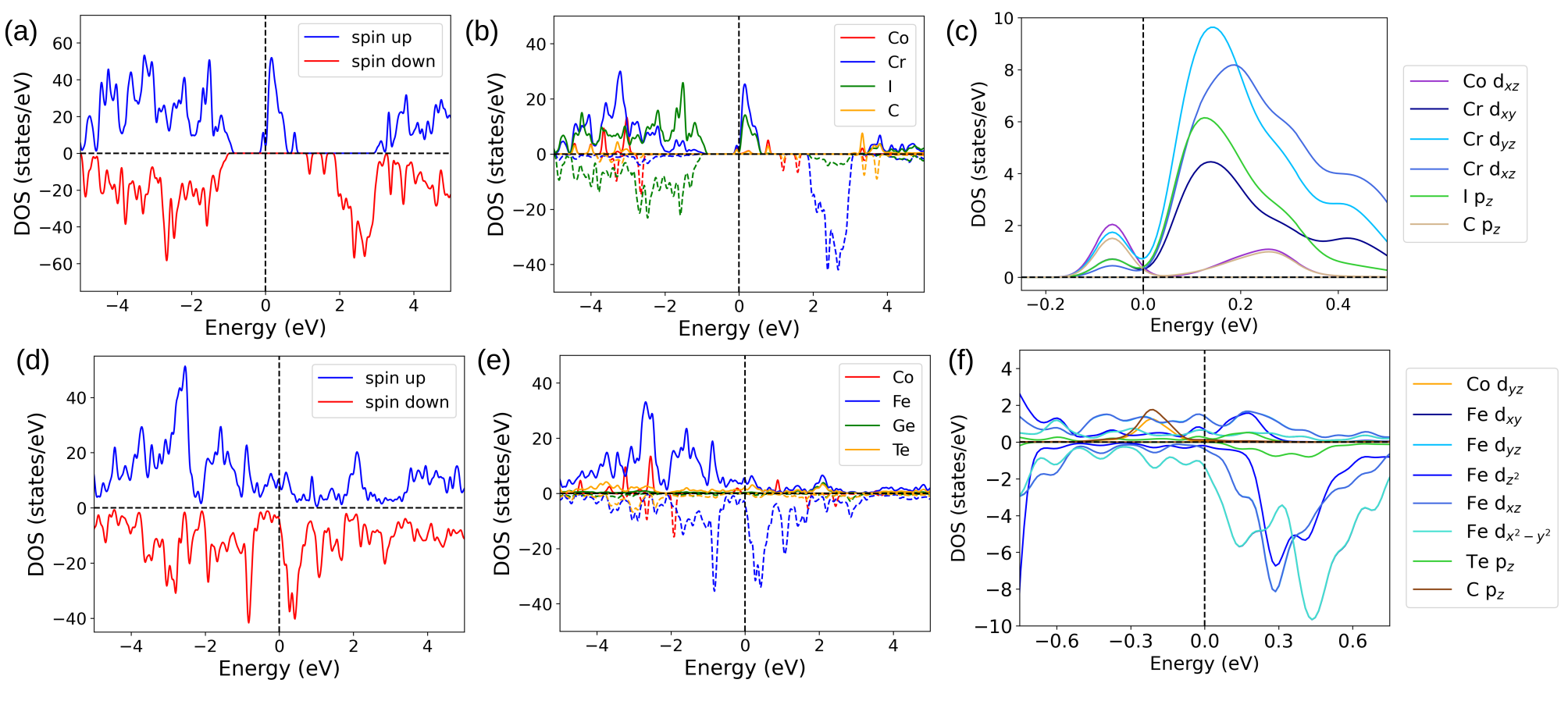}
   \caption{Spin-polarized total (a,d), element-projected (b,e), and orbital-decomposed (c,f) density of states close to the Fermi level for CoCp$_2$/CrI$_3$ (a--c) and CoCp$_2$/Fe$_3$GeTe$_2$ (d--f) interfaces. The (a,b,d,e) plots show a broader interval from \mbox{-5}~eV to 5~eV around the Fermi level ($E_F$ was set to zero), the (c,f) part provides a more detailed view of the states close to the Fermi level.}
    \label{fig3}
\end{figure*}

While freestanding CrI$_3$ is a~semiconductor with a band gap of around 1.1~eV, the Fermi level cuts finite number of states in the spin up channel at the interface (Fig.~\ref{fig3}). %Orbital projections showed that iodine levels dominate in what was previously the top of valence band, similarly to the freestanding CrI$_3$. 

The numerical integration of density of states for 2$\times$2 cell of CrI$_3$ with CoCp$_2$ between \mbox{-0.2~eV} and the Fermi level showed that 0.98 states are present in the given energy range, which is in agreement with the assumption of one unpaired electron in the system. Therefore, the half-metallic nature of the interface seems to originate from the excess electron density of the molecule, in analogy to $n$-type doping.
The electronic states below the Fermi level mainly consist of hybridized Co d$_{xz}$, Cr d$_{yz}$ and C $p_z$ (Fig.~\ref{fig3}c), which suggests significant electronic interaction between cobalt and chromium atoms. This is also evident in the partial charge density for these states (Fig.~\ref{fig4}a).

Moreover, there is a~strong peak in the majority spin channel just above the Fermi level with hybridized Cr $d$ states, mainly d$_{yz}$ and d$_{xz}$, and I $p$ states with smaller contribution of
Co $d$ and C $p$ states. This could open new possibilities for electron doping, effectively
shifting the Fermi level towards the significant peak.

On the other hand, the CoCp$_2$/FGT interface does not show significant changes in the electronic structure in comparison to the~freestanding substrate and remains metallic (Fig.~\ref{fig3}).
The density of states at the Fermi level is higher in the spin-up channel than in the spin-down channel which leads to a~partial spin polarization of the conduction electrons. The reverse is true for energy range of 0.15~eV to 0.6~eV, where the number of available states in the spin-down channel dominates. This finding could be of interest for an~easy manipulation of transport properties in this system. Based on orbital-resolved density of states, the electronic structure at the Fermi level is mostly determined by Fe $d$ states, particularly $d_{z^2}$ and $d_{x^2-y^2}$, with small contribution from Te~$p$ states (Fig.~\ref{fig3}f). The HOMO of the molecule, which mostly consists of Co $d_{yz}$ orbital, exhibits a~broadened peak centered around \mbox{$-$0.25}~eV below the Fermi level. It is shifted in comparison with its position in isolated CoCp$_2$, most probably due to charge redistribution at the interface. 

\subsection{Magnetic properties}
The spin magnetic moments were obtained from GGA+U collinear calculations for both interfaces and compared to the corresponding freestanding substrates.
In the CrI$_3$-based system, Cr atoms show moments between 3.36~$\mu_B$ (Cr$_4$, Cr$_5$) and 3.46~$\mu_B$ (Cr$_6$). Induced moments on iodine atoms are smaller in comparison and range from $-0.13$ to $-0.14$~$\mu_B$. In freestanding CrI$_3$, Cr atoms carry a~moment of 3.36~$\mu_B$, and I atoms have the same moment of $-0.14~\mu_B$. It was found that Co atom carries a~moment of 0.55~$\mu_B$, which is smaller than in the freestanding case, most probably due to charge transfer.

In the FGT system, the Co atom exhibits a~moment of 0.92~$\mu_B$. For Fe atoms, magnetic moments range from 2.43 to 2.45~$\mu_B$ in the first Fe layer, 2.41 to 2.46~$\mu_B$ in the topmost Fe layer (with the largest value on Fe$_7$), and 1.40 to 1.49~$\mu_B$ in the middle Fe layer (maximum moment found on Fe$_9$). The moment on Ge atoms is constant throughout the supercell ($-0.10~\mu_B$), and Te atoms carry even smaller moments around $-0.04~\mu_B$. %Carbon atoms show small induced moments between $-0.02$ and 0.03~$\mu_B$. 
In the pristine FGT structure, magnetic moments were 2.45~$\mu_B$ for the Fe$_1$ and Fe$_3$ sublattices, and 1.44~$\mu_B$ for Fe$_2$, with induced moments of $-0.11~\mu_B$ on Ge and $-0.04~\mu_B$ on Te.

%magnetic ground state
At first, the magnetic ground state of the interfaces was explored at the
GGA+U level. %Both substrates show ferromagnetic order within the layer and freestanding molecule exhibits a~magnetic moment perpendicular to the carbon rings.
The systems were structurally optimized both in FM and AFM surface-molecule coupling and total energies were compared. In all the calculations, the molecular moment was flipped to realize the AFM coupling between the molecule and the magnetic substrate. Our calculations reveal that the ferromagnetic coupling is preferred for both interfaces.

%Exchange interaction
Magnetic exchange interactions contain valuable information about the magnetic order. The magnetic exchange parameters can be obtained via multiple methods. Firstly, a~model Hamiltonian can be used to estimate an~effective interaction from total energy difference between FM and AFM coupling. Another option is to use LKAG formulation employing localized basis, such as an~atomic orbitals basis or maximally localized Wannier functions.

\begin{figure*}[htp]
    \centering
\includegraphics[width=0.95\textwidth]{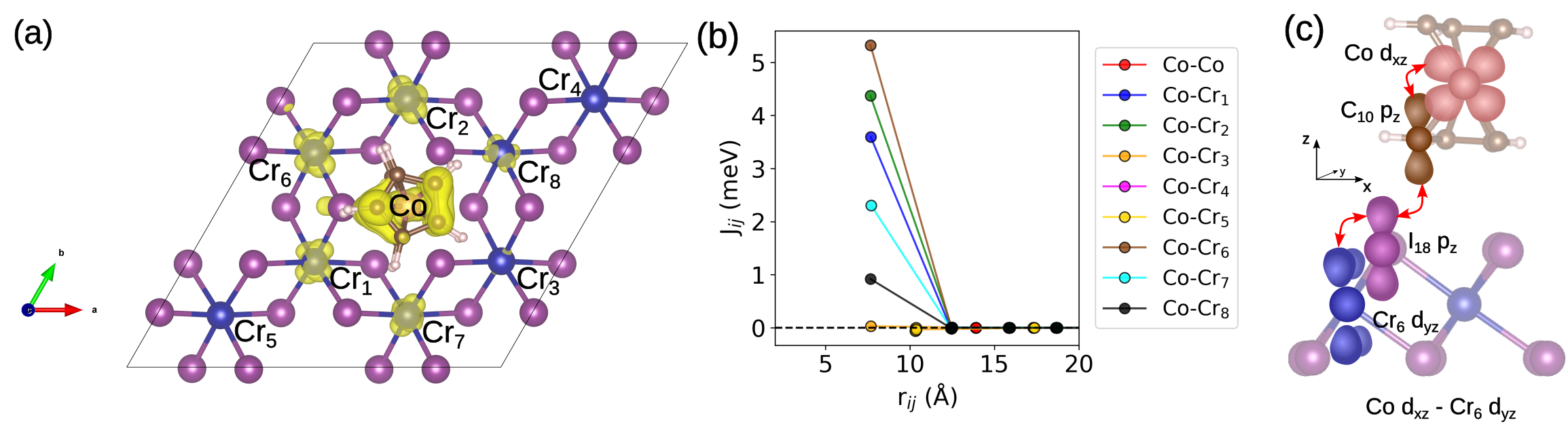}
    \caption{(a): Top view of the partial charge density for states between -0.2 eV and $E_F$ in the CrI$_3$ supercell, with isosurface level $ 10^{-3} \, e$ and Cr atoms labeled. (b): $J_{ij}$ interaction parameters as a function of distance between atoms $i$ and $j$ obtained from TB2J calculation.
    (c): Schematic visualization of an indirect exchange mechanism across the interface, where hoppings are denoted by red arrows.}
    \label{fig4}
\end{figure*}

%rewrite this section
Using the Wannier basis, highly non-degenerate Heisenberg exchange interactions for the Co-Cr nearest-neighbors were found (Fig.~\ref{fig4}) despite having the same interatomic distance. This reflects the broken symmetry due to the molecule having a 5-fold rotation axis while the substrate’s symmetry is 6-fold. The strongest interaction of 5.3~meV is found for Co–Cr$_6$. GGA calculations confirmed non-degenerate nearest-neighbor Co–Cr interactions with smaller $J_{ij}$ magnitude while the order was preserved. This finding excludes artificial symmetry breaking from incorporating the Hubbard correction.

The direct orbital overlap between Co and Cr atoms is negligible and therefore the magnetic interactions are expected to be indirect. In order to gain insight into interaction mechanism, the exchange paths for indirect exchange interaction between Co and the closest Cr atoms with two mediating orbitals of nonmagnetic atoms (Co--C--I--Cr) were analyzed. According to the orbital-decomposed exchange, the strongest contributions to the Co--Cr$_6$ interaction involve Co $d_{xz}$ with Cr$_6$ $d_{yz}$ (2.39~meV), $d_{x^2-y^2}$ (0.72~meV), and $d_{xy}$ (1.70~meV). For the Co $d_{xz}$--Cr$_6$ $d_{yz}$ interaction, the dominant exchange path is likely to be Co $d_{xz}$--C$_{10}$ $p_z$--I$_{18}$ $p_z$--Cr$_6$ $d_{yz}$ (Fig.~\ref{fig4}c) based on hopping parameters.

The exchange interactions were also estimated from total energy differences. The placement of CoCp$_2$ on CrI$_3$ is symmetric. Nearest-neighbor surface interactions were estimated from total energy differences between ferromagnetic (FM) and antiferromagnetic (AFM) configurations. Collinear calculations yield the energy difference $\Delta E=98.84$~meV, which indicates a~strong exchange coupling between the molecule and the surface.

The Co atom is located nearly at the center of a~chromium ring (Fig.~\ref{fig1}) with an average Co–Cr distance of 7.67~\AA. In this particular case, one can consider six nearest-neighbor interactions. Mapping this to an Ising model:
\begin{equation}
E_\text{FM}-E_\text{AFM}=-2\sum_{i,j} J_{ij} S_i S_j ,
\end{equation}
where $i$ is identified with Co and $j$ runs over the nearest-neighbor chromium atoms. Since both $ij$ and $ji$ are counted, an additional factor of 2 appears:
\begin{equation}
E_\text{FM}-E_\text{AFM}=-4\sum_{j} J_\text{eff}\, e_\text{Co} \, e_j ,
\end{equation}
where $J_\text{eff}=J_{ij}S_i S_j$. $J_\text{eff}$ for Co–Cr interaction is obtained as 4.12~meV. %(98.84/24=4.12)

\begin{figure*}[h]
\centering
\includegraphics[width=0.99\textwidth]{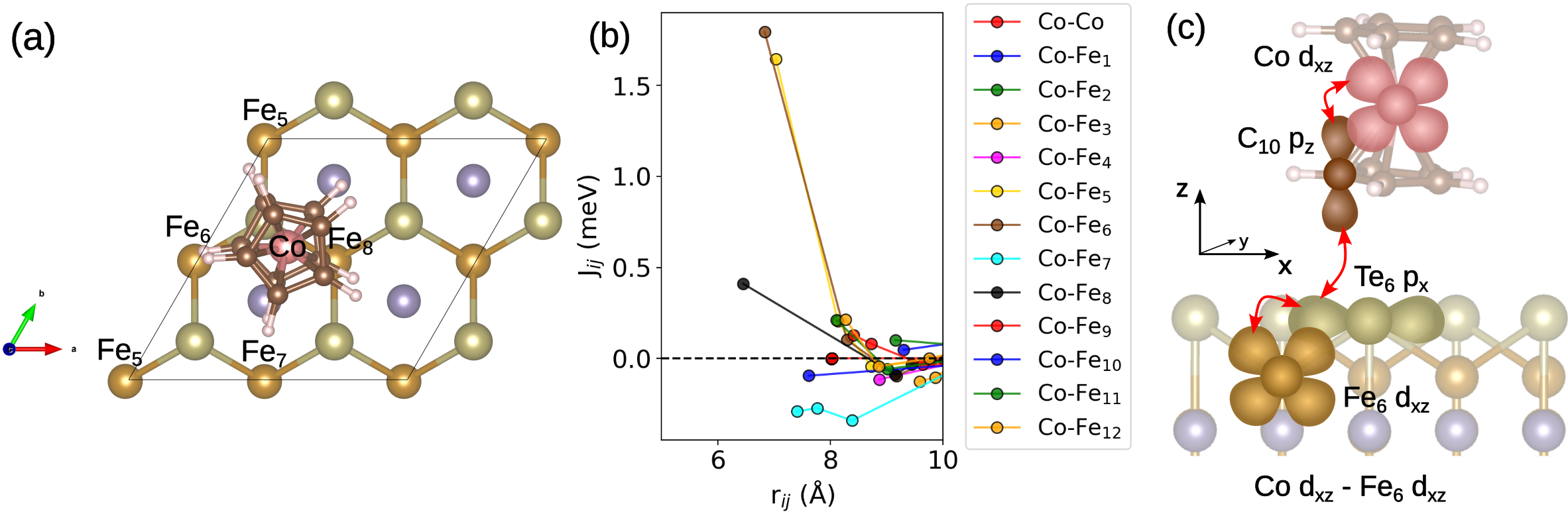}
    \caption{(a): Top view of the FGT supercell with Fe atoms in the upper Fe layer labeled. (b): J$_{ij}$ interaction parameters as a function of distance between atoms $i$ and $j$ obtained from TB2J calculation. (c): Schematic visualization of an indirect exchange mechanism across the interface, where hoppings are denoted by red arrows.}
    \label{fig5}
\end{figure*}

In the FGT supercell, there are 12 iron atoms distributed across three atomic layers: atoms Fe$_{1}$-Fe$_{4}$ are located in the bottom layer (furthest from the molecule), Fe$_{5}$-Fe$_{8}$ are in the topmost layer (closest to the molecule), and Fe$_{9}$-Fe$_{12}$ lie in the middle layer. The interface lacks symmetry and the J$_{ij}$ parameters are highly non-degenerate. The individual exchange interactions between Co and Fe atoms were calculated from maximally localized Wannier functions (Fig.~\ref{fig5}). The strongest interaction was found between Co and Fe$_6$ (1.79~meV) at a~distance of 6.83~\AA, followed by Co--Fe$_5$ (1.64~meV at 7.03~\AA) and Co--Fe$_8$ (0.41~meV at 6.45~\AA). The most negative interaction occurs for the Co--Fe$_7$ pair, with a value of \mbox{-0.34}~meV at 8.39~\AA. Since the molecule is not placed symmetrically on the substrate, the distance between Co and Fe atoms varies. However, as in the CoCp$_2$/CrI$_3$ case, the distance is not the main factor determining the strength of the interaction. The J$_{ij}$ value for the most significant interaction of each Co-Fe pair can be found in the SI (Table~S6).

For Co--Fe$_6$, the dominant contributions arises from Co $d_{xz}$--Fe$_6$ $d_{xz}$ (0.97~meV) with the main exchange path being likely Co $d_{xz}$--C$_{10}$ $p_z$--Te$_6$ $p_x$--Fe$_6$ $d_{xz}$ (Fig.~\ref{fig5}c). 
  
In CoCp$_2$/FGT interface, the energy difference between AFM and FM collinear configurations was significantly smaller than in the previous case, only 9.416~meV per supercell. The sum of individual contributions was performed in order to compare this result with LKAG formalism:
\begin{equation}
    %E_\text{sum}=-4\sum_{j} J_{\text{Co}-j} \,S_\text{Co} \,S_{j},
    E_\text{sum}=-4\sum_{\text{j}} J^\text{{(TB2J)}}_{\text{Co}-j}
    %{\text{Co-j}}
    \,e_\text{Co} \,e_j
    %\text{j}
    ,
    \label{eq:FGT-Jij}
\end{equation}
where $J_{\text{Co-}j}$ corresponds to interactions between cobalt and iron atoms.
The expression in Equation~\ref{eq:FGT-Jij} should correspond to the energy difference $E_\text{FM}-E_\text{AFM}$ of 9.42~meV. Unlike in the previous case, the interface is not symmetric and cannot be properly mapped on a~simple model.
Furthermore, FGT is a~metallic magnet, so the interaction between Co of the molecule and Fe atoms of the substrate might not be properly described by just considering the nearest-neighbor interactions. 

During the summation, the range of the interaction was restricted as a~large number of pairs at longer distances together with computational inaccuracy might lead to higher uncertainty. The summation yielded values of 12.68~meV %(3.17*4) 
(10~\AA{} cutoff distance), and 14.20~meV %( 3.55*4using e * e for magmom
(20~\AA{} cutoff). Both values are slightly higher in comparison with total energy difference.

%intralayer
Modification of intralayer exchange interactions due to molecule adsorption was also analyzed. Pristine CrI$_3$ contains two Cr atoms in the unitcell, which form a~honeycomb lattice similar to the one in graphene or hexagonal boron nitride. For easier comparison between the unitcell and the supercell, the chromium atoms Cr$_1$-Cr$_4$ in the supercell are considered to belong to the first sublattice (A), while Cr$_5$-Cr$_8$ were assigned to the second sublattice (B) (Fig.~\ref{fig4}). 

\begin{figure}[h]
    \centering
\includegraphics[width=0.40\textwidth]{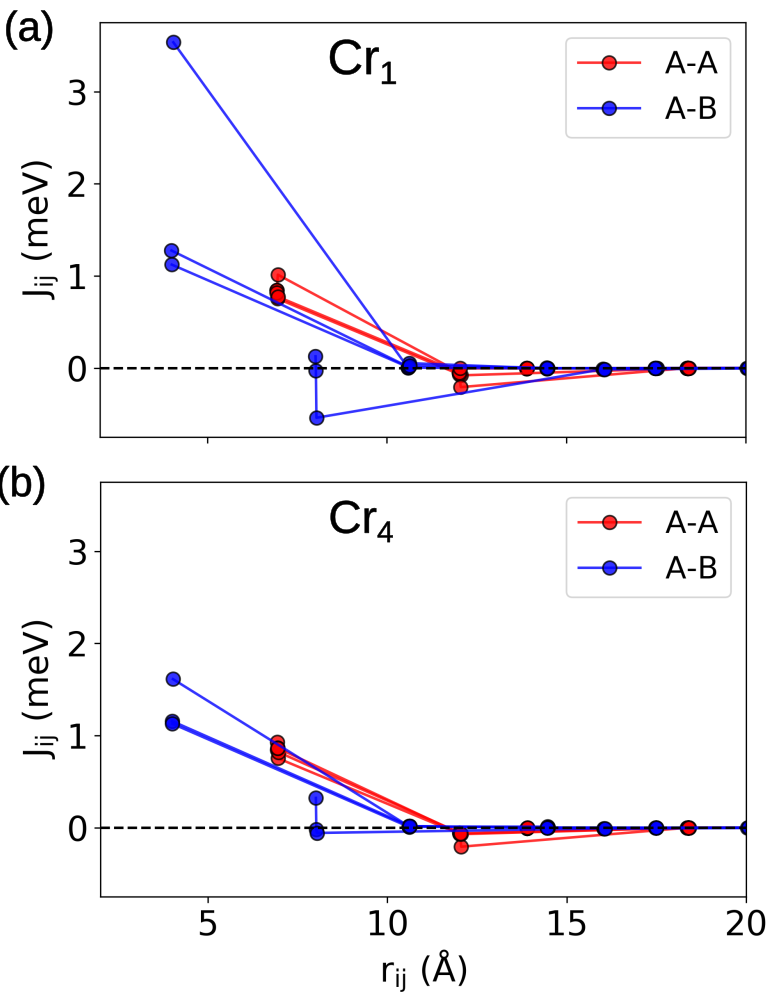}
\caption{Comparison of $J_{ij}$ intralayer interactions from TB2J calculation for Cr$_1$ (a) and Cr$_4$ (b) atoms from the CoCp$_2$/CrI$_3$ structure. The labels $A$ and $B$ in the interaction pairs are based on the two sublattices of Cr. The readers are referred to Fig.~4 for identification of Cr atoms.}
    \label{fig6}
\end{figure}

Threefold symmetry in freestanding CrI$_3$ leads to degenerate nearest-neighbor interactions, but this symmetry is broken upon molecule adsorption. In case of Cr$_1$ atom, two interactions are largely affected (Fig.~\ref{fig6} top): Cr$_1$--Cr$_6$ increases to 3.54~meV, while Cr$_1$--Cr$_8$ becomes significantly negative at \mbox{-0.54}~meV. On the other hand, Cr$_4$, which is located further from the molecule, shows only minor changes in exchange interactions (Fig.~\ref{fig6} bottom). As seen in the partial charge density plot (Fig.~\ref{fig4}a), both Cr$_1$ and Cr$_6$ atoms show strong hybridization with the molecule. Heisenberg exchange parameters for all Cr atoms (Cr$_1$--Cr$_8$) in the supercell can be found in the SI (Fig.~S9).

The mean field estimation of critical temperature for pure CrI$_3$ is 71~K, which increased to 81~K in the presence of the molecule. Metropolis Monte Carlo simulations showed an~increase in T$_\text{C}$ from 45~K for the monolayer to 50~K for the interface obtained from the Binder cumulant crossing method (Fig.~S14), supported by the divergence of susceptibility and heat capacity. We also conclude that by the inspection of intralayer exchange interactions in the FGT monolayer, T$_\text{C}$ is not expected to change significantly.

The magnetocrystalline anisotropy energy (MAE) is defined as the energy difference between the hard and easy magnetization axes. In this study, the MAE is determined as:
\begin{equation}
    E_{\text{MAE}} = E_{\text{in-plane}} - E_{\text{out-of-plane}}.
\end{equation}
Total energy calculations were performed for three distinct orientations of the magnetic moments along the Cartesian axes ($x, y, z$) in $2 \times 2$ supercells. Spin-orbit coupling (SOC) was accounted for as a perturbation applied to the collinear charge density. The symmetry was broken at the CoCp$_2$/CrI$_3$ interface, in contrast to the freestanding CrI$_3$ monolayer, where in-plane orientations are typically degenerate. Consequently, the in-plane MAE values were found to be $4.93$~meV/supercell ($0.62$~meV/f.u.) along the $x$-direction and $1.95$~meV/supercell ($0.24$~meV/f.u.) along the $y$-direction; therefore, the anisotropy is reduced compared to the freestanding case ($0.71$~meV/f.u.). For the CoCp$_2$/FGT system, the MAE is significantly larger than that of the CoCp$_2$/CrI$_3$, yielding values of $13.30$~meV/supercell ($3.33$~meV/f.u.) and $13.27$~meV/supercell ($3.32$~meV/f.u.) for the $x$ and $y$ directions, respectively. Similar to the CrI$_3$-based system, the MAE decreased upon molecular adsorption relative to the freestanding FGT ($3.52$~meV/f.u.).
%Freestanding CrI3 0.705 meV, molecule MAE(x)=77 µeV, and MAE(y)=38 µeV.

\begin{figure}[ht]
    \centering
\includegraphics[width=0.5\textwidth]{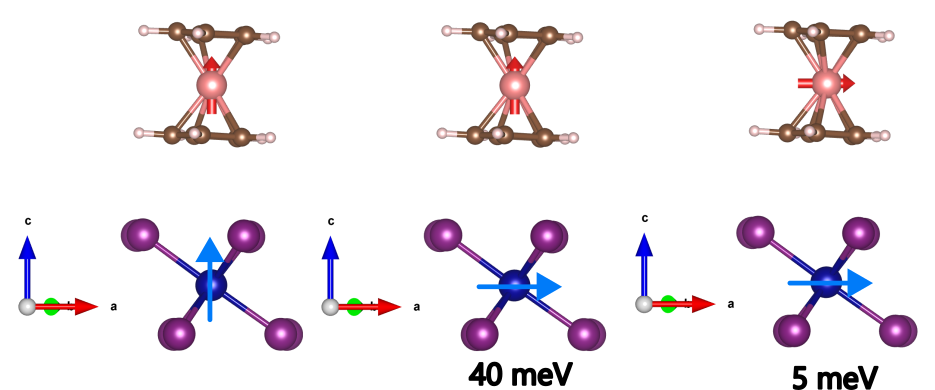}
\caption{Constrained magnetic moment calculations for three different alignments of
magnetic moments in the CoCp$_2$/CrI$_3$. Red and blue arrows indicate the fixed direction of
magnetic moments for the molecule and substrate, respectively. The total energies relative
to the most stable configuration (on the left) are shown below the corresponding structures.}
    \label{fig7}
\end{figure}

%Orbital moments and magnetic anisotropy
To study possible changes in the magnetization behavior due to external factors such as applied magnetic fields, constrained magnetic moment calculations were performed for 3 different alignments of the magnetic moments of the substrate and the molecule: (i) both out-of-plane (Fig.~\ref{fig7} right), (ii) substrate in-plane and molecule out-of-plane (Fig.~\ref{fig7} center), and (iii) both in-plane (Fig.~\ref{fig7} right). All these calculations were done fully self-consistently with the inclusion of spin-orbit coupling. 

The minimum energy was found for out-of-plane orientation of all magnetic moments, in agreement to the MAE calculations presented before, where the SOC effects were treated via the force theorem. The energy difference of 5~meV, when moments were constrained in the Cartesian $x$ and $z$ axes, was close to the value found in MAE calculation (4.93~meV). On the other hand, the highest energy configuration was for the substrate with in-plane magnetization in $x$ direction and out-of plane orientation of molecular magnetic moment with energy difference of approximately 40~meV relative to the ground state. Similar results for CoCp$_2$/FGT can be found in SI (Fig. S13).

%\clearpage
%\newpage

%%%%%%%%%%%%%%%%%%%%%%
\section{Conclusions}
\label{conclusions}

This study explored the interaction between a~magnetic molecule, cobaltocene (CoCp$_2$), and 2D magnetic substrates (CrI$_3$, Fe$_3$GeTe$_2$ monolayers). The main focus was on the determination of structural, electronic, and magnetic properties using first-principles calculations.

%stress characterization of the molecule itself
Initial characterization of substrates was performed for a~reference and showed good agreement with both experimental and theoretical results. Equilibrium geometry and electronic structure for CoCp$_2$ was calculated, including the determination of Hubbard's $U$ to ensure a reliable description in plane-wave basis. The adsorption geometries were obtained by considering various molecule-substrate configurations. The negative adsorption energies showed theoretical stability of the interfaces.

The strength of interaction was determined by the nature of the substrate and orbital hybridization across the interface was observed.
Isotropic exchange parameters ($J_{ij}$) obtained using Green’s function method and total energy differences methods were compared. Strong dependence of $J_{ij}$ parameters on the hybridization of individual atoms and orbitals was obtained via LKAG approach. To further explore modifications of pair exchange interactions, main paths via one or two non-magnetic atoms were analyzed based on hopping parameters from Wannier hamiltonian, considering both interfacial and intralayer exchange. 

Symmetry breaking also became apparent in the MAE calculations, such as significant inequivalence of in-plane directions in CoCp$_2$/CrI$_3$ case. Overall, the preference for out-of-plane magnetization was weakened due to the interfaces.
To evaluate possible response to external stimuli, such as magnetic field, constrained magnetic moment calculations were performed for distinct alignments of magnetic moments of both the substrate and the molecule.
%transport junctions
The CoCp$_2$/CrI$_3$ device showed conducting behavior and the possibility of highly spin-polarized current.  %add about FGT
We envisage that the devices based on these junctions may find applications as part of molecular probes for magnetism, possibly focusing on substrates with out-of-plane easy magnetization axis. Additionally, the results could guide the design of devices suitable for quantum computing or spin transport applications.
%add limitations
%Limitations and uncertainties in the accuracy could be addressed through alternative computational approaches. Using a~localized basis for the electronic structure calculation, e.g., linear muffin-tin orbitals (LMTO), may improve the precision of $J_{ij}$ parameter determination. Additionally, more advanced many-body techniques, such as dynamical mean-field theory (DMFT), could provide better description of electron correlation. %further research?
Future work might explore molecules containing multiple magnetic centers that exhibit complex exchange interactions. Moreover, utilizing substrates with various magnetic anisotropies or modulating their properties via doping or external fields might lead to interfaces with desired properties. Lastly, the role of defects or temperature could also be taken into account to mimic realistic experimental conditions.

%greater context of the results

%%%%%%%%%%%%%%%%%%%%%%
\section{Acknowledgments}
B.S. acknowledges financial support from Swedish Research Council (grant no. 2022-04309) and STINT Mobility Grant for Internationalization (grant no. MG2022-9386). The computational resources provided by the National Academic Infrastructure for Supercomputing in Sweden (NAISS) at UPPMAX (NAISS 2024/5-258) and at NSC and PDC (NAISS 2024/3-40) partially funded by the Swedish Research Council through grant agreement no. 2022-06725.

%\clearpage
%\printbibliography
\bibliography{refs}

\end{document}